\documentclass{elbioimp2}
\usepackage[utf8]{inputenc}

\usepackage[backend=biber,style=vancouver]{biblatex}
\usepackage{csquotes}
  
\usepackage{amsmath,amsfonts}
\usepackage{algorithmic}
\usepackage{graphicx}
\usepackage{textcomp}
\usepackage{booktabs}
\usepackage{url}
\usepackage{multirow}
\usepackage{enumitem}
\usepackage{array, makecell}
\usepackage{hyperref}
\usepackage{float}

\usepackage{subcaption,graphicx}

\title{Explainable Medical Image Segmentation via Generative Adversarial Networks and Layer-wise Relevance Propagation}

\shorttitle{Explainable Medical Image Segmentation via GAN and LRP}
\author{Awadelrahman M. A. Ahmed\affiliation{Department of Informatics, University of Oslo, Norway, aahmed@ifi.uio.no} and
        Leen A. M. Ali\affiliation{Ullevål School, Oslo, Norway}
} 

\shortauthor{Awadelrahman et al.}
\elbioimpreceived{1 Jan 2021}  
\elbioimppublished{21 Jul 2021}
\elbioimpfirstpage{1}
\elbioimpvolume{10}
\elbioimpyear{2021}

\begin{document}
\maketitle

\begin{abstract}
This paper contributes to automating medical image segmentation by proposing generative adversarial network-based models to segment both polyps and instruments in endoscopy images. A major contribution of this work is to provide explanations for the predictions using a layer-wise relevance propagation approach designating which input image pixels are relevant to the predictions and to what extent. On the polyp segmentation task, the models achieved 0.84 of accuracy and 0.46 on Jaccard index. On the instrument segmentation task, the models achieved 0.96 of accuracy and 0.70 on Jaccard index.  The code is available at https://github.com/Awadelrahman/MedAI.
\end{abstract}

\section{Introduction}
A polyp is a small assembly of cells that forms on the lining of the colon. Even though colon polyps are mostly harmless, over time, some colon polyps may develop into cancer. Automating medical image segmentation can reduce the percentage of overlooked polyps. However, with the recent increase in using machine learning approaches in the field, it is becoming increasingly difficult for experts to interpret the decisions made by black-box models. Therefore, interpretability becomes as important as prediction accuracy; thus, transparent and explainable machine-learning models are required to answer the "why" and "how" questions in addition to producing accurate predictions. Motivated by that, this paper presents a submission to \emph{MedAI: Transparency in Medical Image Segmentation Challenge}~\cite{MediAI2021} which aims to develop transparent and explainable automatic segmentation systems.

\begin{figure}[htpb]
    \centering
    \includegraphics[width=0.87\columnwidth]{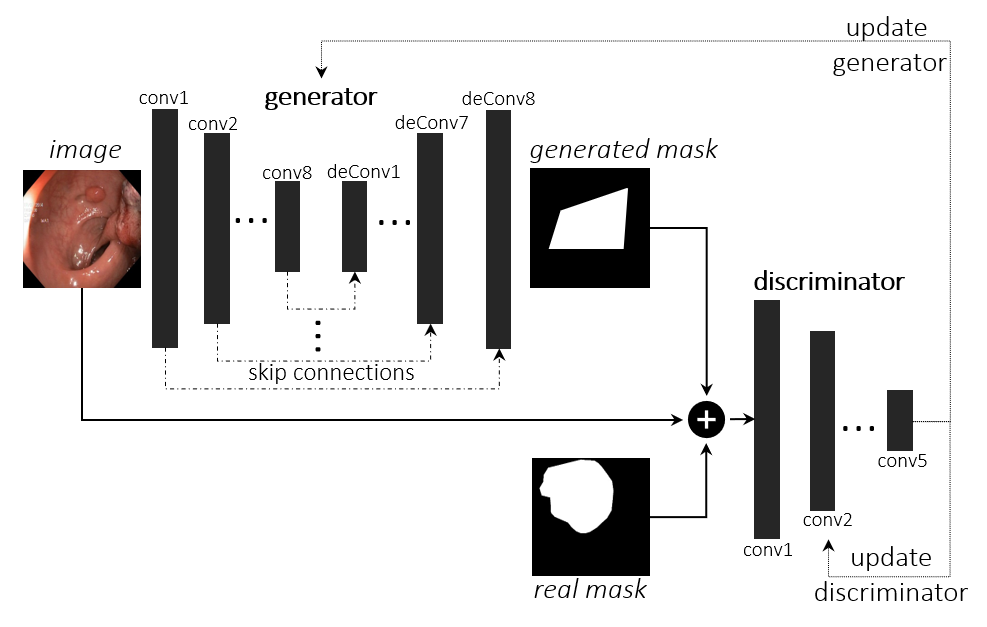}
    \caption{Model architecture. The + represents a concatenation operation (see Table~\ref{arch}\label{bd}).}  
    \vspace{-5mm}
\end{figure}

Given gastrointestinal polyp images, to produce corresponding masks that locate the polyps or instruments, we utilize the generative adversarial networks (GANs) framework that has been successfully implemented to solve similar image-to-image translation problems~\cite{isola2017image}. To characterize our models transparency and explainability, we adopt the layer-wise relevance propagation (LRP) approach~\cite{bach2015pixel}, one of the most prominent methods in explainable artificial intelligence (XAI). We generate relevance maps that show the contribution of each pixel of the input image in the final decision of the model. The rest of the paper illustrates the methods and discusses the evaluation results.

\section{GANs for Polyp and Instrument Segmentation}
The GAN-based segmentation model consists of a bottle-neck like \textit{generator} network which has a convolution-layers encoding part and a de-convolution-layers decoding part with skip connections between them. The generator takes the images as input and produces conditioned realistic-looking masks. A \textit{discriminator} which a CNN-based network that has access to the ground truth masks and classifies whether the generated masks are real or not. The masks are concatenated with their input images before being fed to the discriminator. Both networks are trained adversarially based on the discriminator output. After a finite number of epochs, the generator will be able to generated masks that are hardly classified by the discriminator as fake. The model block diagram is shown in Figure~\ref{bd} and the model details are shown in Table~\ref{arch}. We use the same architecture for the polyp segmentation task and the instrument segmentation.

		
		

\begin{table}[h]
\caption{Model details}\label{arch}
\begin{tabular}{|c|l|c|c|}
\cline{2-4}
\multicolumn{1}{c|}{} & \multicolumn{1}{c|}{Layer} & Filters num. & Activation \\ \hline
\parbox[t]{1mm}{\multirow{9}{*}{\rotatebox[origin=c]{90}{Generator}}}  & Conv1 &  $64\times 1$  & LeakyReLU \\ \cline{2-4} 
                             & Conv2 &  $64\times 2$ & LeakyReLU  \\ \cline{2-4} 
                             & Conv3 &  $64\times 4$ & LeakyReLU\\ \cline{2-4} 
                             & Conv [4 - 8] &  $64\times 8$ & LeakyReLU\\ \cline{2-4}
                             & DeConv [1 - 4] &  $64\times 8$ & ReLU \\ \cline{2-4}
                             & DeConv5 &  $64\times 4$  & \ ReLU  \\ \cline{2-4}
                             & DeConv6 &  $64\times 2$ & \ ReLU \\ \cline{2-4}
                             & DeConv7 &  $64\times 1$ & \ ReLU \\  \cline{2-4}
                             & DeConv8 &  1& \ Tanh     
                             \\ \hline
\parbox[t]{2mm}{\multirow{5}{*}{\rotatebox[origin=c]{90}{Discriminator}}} &  Conv1 &  $64 \times 1$ &   LeakyReLU     \\  \cline{2-4} 
                             &	Conv2 &  $64\times 2$ &  LeakyReLU     \\ \cline{2-4} 
                        	&	Conv3 &  $64\times 4$ & \ LeakyReLU     \\ \cline{2-4}
                        	&	Conv4 &  $64\times 8$ & \ LeakyReLU     \\ \cline{2-4}
                        	&	Conv5 &  1 & \ Sigmoid      
                             \\ \hline
                             
\end{tabular}
\end{table}

\section{LRP for Explainable GAN}
The goal of LRP is to explain the neural network’s output in the domain of its input. LRP produces maps that indicate which pixels in the gastrointestinal image contribute to the generated mask and to what extent. In our GAN model, we apply the LRP on the trained generator network since it is the part that is responsible for generating the masks. To calculate a relevance score $Rj$ for every neuron $j$, we start at the output end and pick a neuron that we want to explain. The relevance of this particular neuron is equal to its activation, while the relevance of all other neurons in the output layer are set to zero. From the output as an initial point, we move backwards through the network following the rule in Equation~\ref{lrp}. This rule calculates the relevance $R$ of neuron $j$ in layer $l$ given the relevance scores of every neuron $k$ in layer $l+1$. The fraction calculates the activation of neuron $j$ compared to all activations of other neurons in layer $l$ that will contribute to layer $l+1$. Intuitively, if this relative activation is high, then the specific neuron is highly important for the output. Then we multiply this fraction by the relevance score from the $l+1$ layer to propagate the relevance scores backwards. For efficient implementation of LRP, we followed the implementation guidelines in \cite{montavon2019layer}.

\begin{equation}
	\label{lrp}
    R_j^{(l)}=\sum_k \frac{a_j^{(l)} w_{jk}^{(l,l+1)}}{\sum_{j'} a_{j'} w_{j'k}} R_k^{(l+1)}
\end{equation}

\section{Experimental Evaluation Results and Discussion}
\begin{figure}[htpb]
    \centering
    \begin{subfigure}[b]{0.48\linewidth}
        \includegraphics[width=\linewidth]{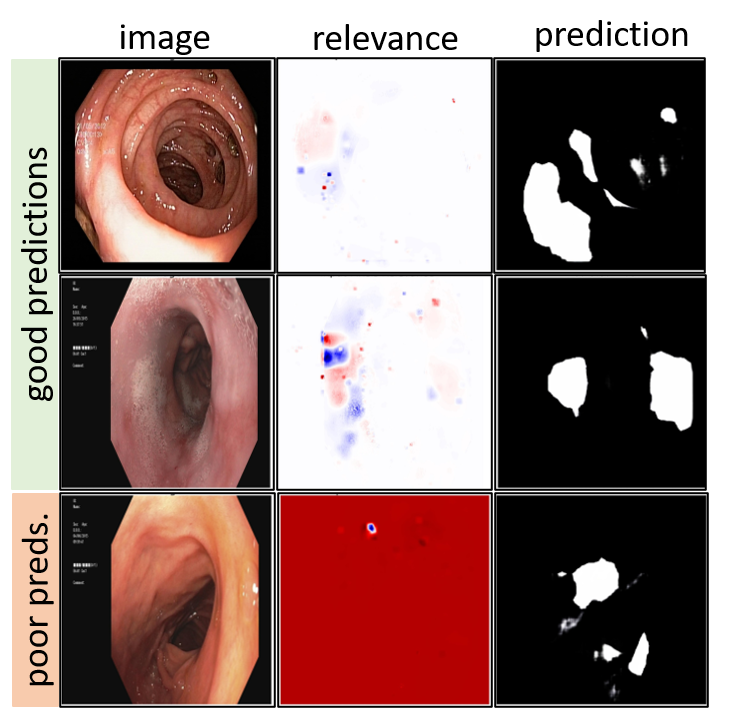}
        \caption{Polyps}\label{rlp1}
    \end{subfigure}
    \begin{subfigure}[b]{.48\linewidth}
        \includegraphics[width=\linewidth]{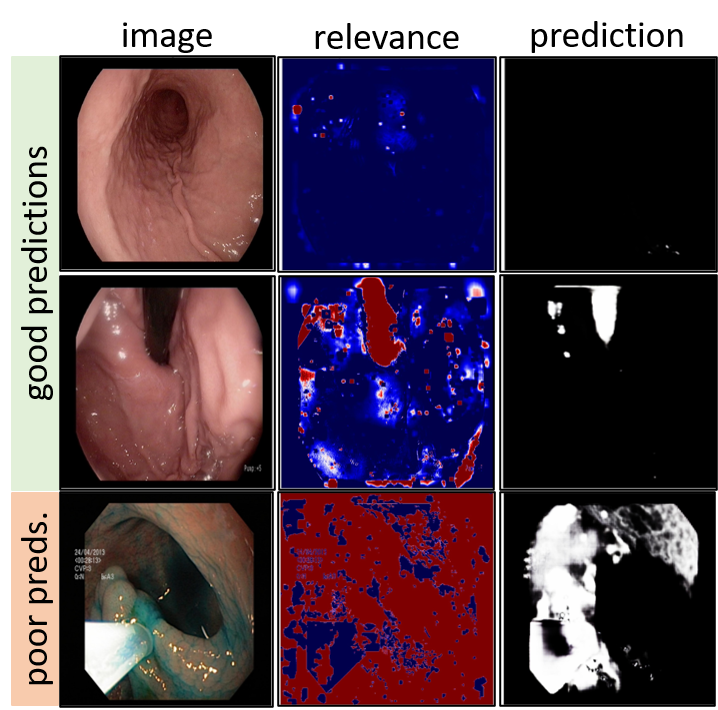}
        \caption{Instruments}\label{rpl2}
    \end{subfigure}

    \begin{subfigure}[b]{0.95 \linewidth}
        \includegraphics[width=\linewidth]{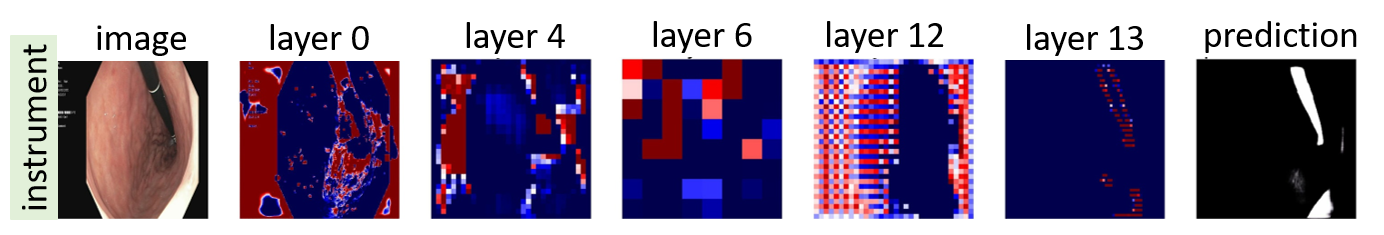}
        \caption{Intermediate layers neurons relevance maps.}\label{rlp3}
    \end{subfigure}
    
    \caption{Predicted masks and relevance maps examples.}\label{rlpfig}
\end{figure}

The GAN-based models and the LRP method were implemented using Python 3.8.5 and the PyTorch 1.9.0 on a 2.5GHz, Intel i5 CPU. Details about the challenge are in \cite{MediAI2021}. We restricted the models training on the polyp data set \cite{jha2020kvasir} and the instrument data set \cite{Jha2020}. Table~\ref{polypres} shows the performance metrics on both tasks' official test set. The models show higher performance on instruments than on polyps. That is due to the similarity between the polyps' pixels and the normal mucosa pixels in the image, while instruments pixels dissimilarity is higher, creating sharper edges around the instrument pixels and the sudden changes in the pixels values make it easier to be detected by the convolution operations. The input layer relevance maps are shown in Figure~\ref{rlpfig}, the red areas show high relevance and blue areas show low relevance. The high contrast in the instrument maps in Figure~\ref{rpl2} reflects the confidence of the models to decide the pixels' relevance. The models were also able to produce correct masks when there are no instruments present in the image. Figure~\ref{rlp3} shows the relevance maps of intermediate layers and how a particular mask is generated.

\begin{table}[htpb]
    \centering
    \caption{Models results on the official provided test set.}\label{polypres}
    \begin{tabular}{ |c|c|c|c|c|c| }
        \hline
        Task &	Acc. & Jaccard & Dice & Recall &  Precision \\ \hline
        Polyp & 0.84 & 0.46 & 0.56 & 0.64 & 0.59 \\ \hline
        Inst. & 0.96 & 0.70 & 0.80 & 0.80 & 0.80 \\ \hline
    \end{tabular}
    \vspace{-5mm}
\end{table}

\subsection{Conflict of interest}
Authors state no conflict of interest. 

\newpage
\printbibliography

\end{document}